\documentclass[pra,twocolumn,superscriptaddress,10pt]{revtex4-1}
\usepackage{graphicx}
\usepackage{dcolumn}
\usepackage{color}
\usepackage{times}
\usepackage{bm}
\usepackage{amssymb}
\usepackage{amsmath}
\usepackage{epsfig}
\usepackage{epstopdf}
\usepackage{dsfont}
\usepackage{subfigure}
\begin{document}
\renewcommand{\thefootnote}{\fnsymbol {footnote}}
	
	\title{\textbf{Constraint relation between steerability and concurrence for two-qubit states}}
	
	\author{Xiao-Gang Fan}
	\affiliation{School of Physics \& Material Science, Anhui University, Hefei 230601, China}
	
	\author{Huan Yang}
	\affiliation{School of Physics \& Material Science, Anhui University, Hefei 230601, China}
	\affiliation{Department of Experiment and Practical Training Management, West Anhui University, Lu'an 237012, China}
	
	\author{Fei Ming}
	\affiliation{School of Physics \& Material Science, Anhui University, Hefei 230601, China}

    \author{Zhi-Yong Ding}
	\affiliation{School of Physics \& Material Science, Anhui University, Hefei 230601, China}
	\affiliation{School of Physics and Electronic Engineering, Fuyang Normal University, Fuyang 236037, China}

	\author{Dong Wang}
	\affiliation{School of Physics \& Material Science, Anhui University, Hefei 230601, China}
	
	\author{Liu Ye}
	\email{yeliu@ahu.edu.cn}
	\affiliation{School of Physics \& Material Science, Anhui University, Hefei 230601, China}
	
\begin{abstract}
Entanglement and steering are used to describe quantum inseparabilities.  Steerable states form a strict subset of entangled states. A natural question arises concerning how much territory steerability occupies entanglement for a general two-qubit entangled state. In this work, we investigate the constraint relation between steerability and concurrence by using two kinds of evolutionary states and randomly generated two-qubit states. By combining the theoretical and numerical proofs, we obtain the upper and lower boundaries of steerability. And the lower boundary can be used as a sufficient criterion for steering detection. Futhermore, we consider a special kind of mixed state transformed by performing an arbitrary unitary operation on Werner-like state,  and propose a sufficient steering criterion described by concurrence and purity.
\end{abstract}
\maketitle

\section{Introduction}
In 1935,  Schr\"odinger   initially introduced the concept of steering as a generalization of Einstein-Podolsky-Rosen (EPR) paradox \cite{S1}.  It is supposed that Alice and Bob share an entangled state in two distant area. Alice can steer the particle of Bob into different states by performing measurements on her own paticle.  Recently, it has  been formalized that steering is a sufficient form of quantum inseparabilities \cite{S2}. And quantum inseparability is usually measured by entanglement.

In fact, steerable states form a strict subset of entangled states, and this means that not every entangled state is steerable \cite{S2}. This steerability is manifested explicitly by violating the different steering inequalities, and it plays an essential role in better understanding  the subtle aspects of quantum mechanics. Historically, the violation of steering inequality is usually used as a criterion for whether two-qubit pure states are entangled. However, for a two-qubit mixed state, it is not applicable in this case. Wiseman \textit{et al.}  demonstrated that Werner state with weak entanglement does not violate any steering inequalities \cite{S2}. Afterwards, it is  experimentally proved  that EPR-steering is captured for mixed entangled states that are Bell local \cite{S3}. Later on, Costa \textit{et al.} proposed a measure of steering that is based on the maximal violation of the line steering inequality in the two- and three-measurement scenarios \cite{S4}. Recently, Guo \textit{et al.} \cite{S6} presented a steering criterion which is both necessary and sufficient for two-qubit states under arbitrary measurement sets. And   a set of complementarity relations between  steering or  nonlocal advantage of quantum coherence inequalities can be derived and achieved by various criteria \cite{S5}.  Das \textit{et al.} \cite{B2} proposed a criterion to detect whether a given two-qubit state is steerable. To be specific,  for  a  target state, the new state is constructed. The steering of target state can be detected via detecting  entanglement of the new state \cite{B2,yang}.

Now that both entanglement and steering are used to describe quantum inseparabilities. As two vital quantum resources, we are more concerned about how much domain the values of steerability are limited in the values of entanglement. And what states do the  boundary states of steerability located in entanglement region represents? In addition, we aim to explore the constrained relation between two quantum resources  for arbitrary two-qubit states.

The remainder of this paper is organized as follows. In Sec. II, we review the related concepts of concurrence and steering. In Sec. III, we investigate the relation between concurrence and steerability by using the evolutionary states (Bell-like state under two kinds of deconherence channels) and randomly generated two-qubit states. Furthermore, it is obtained that  the upper and lower boundaries of steerability  can be expressed by concurrence and purity.   In Sec. IV, we investigate the relation between concurrence and steering for a special kind of mixed states (the states are transformed by an arbitrary unitary operation on Werner-like states). In final, we end up our article with a brief conclusion.

\section{Preliminaries} \label{sec2}
Concurrence is usually used as a measure for entanglement of two-qubit states \cite{w01,w25}. For a two-qubit pure state $\left| \psi  \right\rangle $, its concurrence is defined as \cite{w26}
\begin{align}
C\left( {\left| \psi  \right\rangle } \right) = \left| {\left\langle {\psi }
	\mathrel{\left | {\vphantom {\psi  {\tilde \psi }}}
		\right. \kern-\nulldelimiterspace}
	{{\tilde \psi }} \right\rangle } \right|,
\end{align}
where $\left| {\tilde \psi } \right\rangle {\rm{ = (}}{\sigma _y} \otimes {\sigma _y})\left| {{\psi ^*}} \right\rangle $. Here $\left| {{\psi ^*}} \right\rangle $ is the complex conjugate of the pure state $\left| \psi  \right\rangle $ and ${\sigma _y}$ is the Pauli-y matrice. For a general two-qubit state $\rho $, its concurrence is defined by the convex-roof \cite{w11,A} 
\begin{align}
C\left( \rho  \right) = \mathop {\min }\limits_{\left\{ {{q_n},\left| {{\varphi _n}} \right\rangle } \right\}} \sum\limits_n {{q_n}C\left( {\left| {{\varphi _n}} \right\rangle } \right)}.
\end{align}
The minimization is taken over all possible decompositions $\rho $ into pure states.
An analytic solution of concurrence can be calculated \cite{w26}
\begin{align}
C\left( \rho  \right) = \max \left\{ {0,2\sqrt {{\lambda _1}\left( \rho  \right)}  - \sum\limits_{n = 1}^4 {\sqrt {{\lambda _n}\left( \rho  \right)} } } \right\},
\end{align}
where ${\lambda _n\left( \rho  \right)}$ are the eigenvalues, in decreasing order, of the non-Hermitian matrix $\rho \tilde \rho $. Here, the matrix $\tilde \rho $ which is the spin-flipped density matrix of the state $\rho$, has the following form
\begin{align}
\tilde \rho  = {\rm{(}}{\sigma _y} \otimes {\sigma _y}){\rho ^*}{\rm{(}}{\sigma _y} \otimes {\sigma _y}),
\end{align}
where the matrix ${\rho ^*}$ is the complex conjugate of the state $\rho $.

Besides, steering inequality violation in quantum mechanics clearly illuminates that quantum correlations are quite different from classical correlations. In the case of two-qubit states, Cavalcanti-Jones-Wiseman-Reid (CJWR) inequality \cite{S4} is a well-known steering inequality. And it has the important property that an arbitrary two-qubit pure state may violate CJWR inequality if this state is entangled. The violation of CJWR inequality diagnoses whether a two-qubit state is steerable when Alice and Bob are both allowed to measure $N$ observables in their sites. More recently, Costa \textit{et al.}  gave the analytical results of $N=2,3$ for any two-qubit states \cite{S4}. Setting $N=3$,  the inequality decribed by a finite sum of bilinear expectation values is
\begin{align}
F^{{\rm{CJWR}}}\left( {\rho ,r} \right) = \frac{1}{{\sqrt 3 }}\left| {\sum\limits_{i = 1}^3 {\left\langle {{A_i} \otimes {B_i}} \right\rangle } } \right| \le 1,
\end{align}
where ${A_i} = \hat r_i^{\rm{A}} \cdot \vec \sigma $, ${B_i} = \hat r_i^{\rm{B}} \cdot \vec \sigma $ are used to be describe the projection measurements on sides A and B, respectively. Here, $\vec \sigma  = \left( {{\sigma _x},{\rm{ }}{\sigma _y},{\rm{ }}{\sigma _z}} \right)$ is a vector made up of Pauli matrices, $r = \left\{ {\hat r_1^{\rm{A}},\hat r_2^{\rm{A}}, \hat r_3^{\rm{A}};\hat r_1^{\rm{B}},\hat r_2^{\rm{B}}, \hat r_3^{\rm{B}}} \right\}$ is the set of measurement directions, and $\left\langle {{A_i} \otimes {B_i}} \right\rangle  = {\rm{Tr}}\left( {\rho {A_i} \otimes {B_i}} \right)$ is the expected value of the projection operator ${A_i} \otimes {B_i}$ in the state $\rho$. Considering maximally values of $F^{{\rm{CJWR}}}\left( {\rho ,r} \right)$, Eq. (5)  can be rewritten as
\begin{align}
F\left( \rho  \right) = \sqrt {t_1^2\left( \rho  \right) + t_2^2\left( \rho  \right) + t_3^2\left( \rho  \right)}\le 1,
\end{align}
where ${t_i}\left( \rho  \right){\rm{  }}\left( {i \in \left\{ {1,2,3} \right\}} \right)$ are the eigenvalues, in decreasing order, of the matrix  $t\left( \rho  \right) = \sqrt {{T^T}\left( \rho  \right)T\left( \rho  \right)}$ with a correlation matrix ${T(\rho)}$ and the transpose matrix ${T^T }\left( \rho  \right)$. The real matrix ${T(\rho)}$ is formed by the coefficients $ {\rm Tr}\left( {\rho {\sigma _m} \otimes {\sigma _n}} \right)$ ($m,{\rm{ }}n \in \left\{ {x,{\rm{ }}y,{\rm{ }}z} \right\}$). Here the quantity $F\left( \rho  \right)$ donates  a three-measurement bilinear maximally expectation values. In order to render steerability $0 \le S\left( \rho  \right) \le 1$, we consider steerability has the following form
\begin{align}
S\left( \rho  \right) = \sqrt {\frac{1}{2}\max \left\{ {0,{F^2}\left( \rho  \right) - 1} \right\}}.
\end{align}
\section{Constraint relation between steerability and concurrence} \label{sec3}
Following these Refs. \cite{w27,A}, we can obtain that any pure states ${\left| \varphi  \right\rangle }$ satisfy two kinds of complementary equations, i.e.,
\begin{align}
\frac{{1 + 2{D^2}\left( {\left| \varphi  \right\rangle } \right) + {F^2}\left( {\left| \varphi  \right\rangle } \right)}}{4} = 1,
\end{align}
\begin{align}
{C^2}\left( {\left| \varphi  \right\rangle } \right) + {D^2}\left( {\left| \varphi  \right\rangle } \right) = 1,
\end{align}
where ${D\left( {\left| \varphi  \right\rangle } \right)}$ is first-order coherence of the pure states ${\left| \varphi  \right\rangle }$. Combining Eqs. (8) and (9),  the quantity $F\left( \left| \psi  \right\rangle  \right)$ can be given by
\begin{align}
F\left( {\left| \varphi  \right\rangle } \right) = \sqrt {1 + 2{C^2}\left( {\left| \varphi  \right\rangle } \right)} .
\end{align}
Therefore, for the pure state $\left| \varphi  \right\rangle$, the relation between concurrence $C\left( {\left| \varphi  \right\rangle } \right)$ and steerability $S\left( {\left| \varphi  \right\rangle } \right)$  is 
\begin{align}
S\left( {\left| \varphi  \right\rangle } \right)=C\left( {\left| \varphi  \right\rangle } \right).
\end{align}

It shows that steerability is equivalent to concurrence in a pure state system in Eq. (11) . However, for a general two-qubit mixed state $\rho$, the relation between concurrence $C\left( \rho  \right)$ and steerability  $S\left( \rho  \right)$ is intricate. It is well known that steerable states form a strict subset of entangled states. In other words, steerable states must be entangled states, but entangled states are not necessarily steerable states.  For more clearly investigating the relation between steering and concurrence, we consider the cases for the evolutionary states of Bell-like state going through the amplitude damping (AD) and phase damping (PD) channels, respectively. And we obtain that steerability  can be expressed via concurrence and purity. For any two-qubit states, we give out a constraint inequality relation between two quantum resources and verify it by using lots of randomly generated two-qubit states. 

\subsection{Evolutionary state corresponding to the AD channel}

We consider the output state ${\rho_{\rm{BAD}}}$ which is formed by the particle (A or B) of Bell-like state $\left| {{\varphi _{\rm B}}} \right\rangle$ going through the AD channel. And the state ${\rho_{\rm{BAD}}}$ has the following concise form
\begin{align}
\rho_{\rm{BAD}}  =&\sum\limits_{i = 0}^1 {{K_i}\left| {{\varphi _{\rm B}}} \right\rangle \left\langle {{\varphi _{\rm B}}} \right|K_i^\dag  },
\end{align}
where ${K_0} = \left| 0 \right\rangle \left\langle 0 \right|{\rm{ + }}\sqrt {1 - \eta } \left| 1 \right\rangle \left\langle 1 \right|$ and ${K_1} =\sqrt \eta  \left| 0 \right\rangle \left\langle 1 \right|$ are the Kraus operators of AD channel.
 From Eq. (4), one obtain that the non-Hermitian matrix ${{\rho _{{\rm{BAD}}}}{{\tilde \rho }_{{\rm{BAD}}}}}$  is a matrix of rank 1, i.e.,
\begin{align}
R\left( {{\rho _{{\rm{BAD}}}}{{\tilde \rho }_{{\rm{BAD}}}}} \right) = 1.
\end{align}
Obviously, concurrence $C\left( {{\rho _{{\rm{BAD}}}}} \right)$ and first-order coherence $D\left( {{\rho _{{\rm{BAD}}}}} \right)$ satisify the following relation \cite{A}
\begin{align}
{C^2}\left( {{\rho _{{\rm{BAD}}}}} \right) + {D^2}\left( {{\rho _{{\rm{BAD}}}}} \right) = {\rm{Tr}}\left( {\rho _{{\rm{BAD}}}^2} \right).
\end{align}
By  some calculations,  concurrence of the state $\rho_{\rm{BAD}} $ can be obtained 
\begin{align}
C\left( {{\rho _{{\rm{BAD}}}}} \right){\rm{ }} = \sqrt {{\lambda _1}\left( {{\rho _{{\rm{BAD}}}}} \right)} {\rm{ }} = \sqrt {1{\rm{ - }}\eta } C\left( {\left| {{\varphi _{\rm{B}}}} \right\rangle } \right).
\end{align}
Following the Ref. \cite{w27}, the state $\rho_{\rm{BAD}} $ satisfies the complementary equation, i.e.,
\begin{align}
\frac{{1 + 2{D^2}\left( \rho_{\rm{BAD}} \right) + {F^2}\left( \rho_{\rm{BAD}} \right)}}{4} =  {\rm{Tr}}\left( {\rho _{{\rm{BAD}}}^2} \right).
\end{align}
Combining Eqs. (14) and (16), the quantity $F\left(\rho_{\rm{BAD}} \right)$ for state $\rho_{\rm{BAD}}$ can be expressed as
\begin{align}
F\left( \rho_{\rm{BAD}}  \right) =\sqrt {2{C^2}\left( {{\rho _{{\rm{BAD}}}}} \right) + 2{\rm{Tr}}\left( {\rho _{{\rm{BAD}}}^2} \right) - 1}.
\end{align}
Therefore, steerability $S\left( \rho_{\rm{BAD}}  \right)$ can be expressed in terms concurrence $C\left( \rho_{\rm{BAD}}  \right)$ and purity ${\rm{Tr}}\left( {\rho _{{\rm{BAD}}}^2} \right)$, i.e.,
\begin{align}
S\left( \rho_{\rm{BAD}} \right) = \sqrt {\max \left\{ {0,{Q^2}\left( {{\rho _{{\rm{BAD}}}}} \right) - 1} \right\}},
\end{align}
where $Q\left( {{\rho _{{\rm{BAD}}}}} \right) = \sqrt {{C^2}\left( {{\rho _{{\rm{BAD}}}}} \right) + {\rm{Tr}}\left( {\rho _{{\rm{BAD}}}^2} \right)}$.
It shows that when cocurrence and purity meet the condition ${C^2}\left( {{\rho _{{\rm{BAD}}}}} \right) + {\rm{Tr}}\left( {\rho _{{\rm{BAD}}}^2} \right)>1$,  steerability of the state $\rho_{\rm{BAD}}$ can be deteced. 

\subsection{Evolutionary state corresponding to the PD channel}

We consider the output state ${\rho_{\rm{BPD}}}$ which is formed by the particle (A or B) of Bell-like state $\left| {{\varphi _{\rm B}}} \right\rangle$ going through the PD channel. And the state ${\rho_{\rm{BPD}}}$ has the following concise form
\begin{align}
\rho_{\rm{BPD}}  =&\sum\limits_{i = 0}^1 {{K_i}\left| {{\varphi _{\rm B}}} \right\rangle \left\langle {{\varphi _{\rm B}}} \right|K_i^\dag  },
\end{align}
where ${K_0} = \left| 0 \right\rangle \left\langle 0 \right|{\rm{ + }}\sqrt {1 - \eta } \left| 1 \right\rangle \left\langle 1 \right|$ and ${K_1} =\sqrt \eta  \left| 1 \right\rangle \left\langle 1 \right|$ are the Kraus operators of PD channel. Based on Eq. (4), we can obtain that the non-Hermitian matrix ${{\rho _{{\rm{BPD}}}}{{\tilde \rho }_{{\rm{BPD}}}}}$  is a matrix of rank 2, i.e.,
\begin{align}
R\left( {{\rho _{{\rm{BPD}}}}{{\tilde \rho }_{{\rm{BPD}}}}} \right) = 2.
\end{align}
It reveals that concurrence $C\left( {{\rho _{{\rm{BPD}}}}} \right)$ and first-order coherence $D\left( {{\rho _{{\rm{BPD}}}}} \right)$ satisify the following relation \cite{A}
\begin{align}
{C^2}\left( {{\rho _{{\rm{BPD}}}}} \right) + {D^2}\left( {{\rho _{{\rm{BPD}}}}} \right) \le {\rm{Tr}}\left( {\rho _{{\rm{BPD}}}^2} \right).
\end{align}
For calculating concurrence of the state $\rho_{\rm{BPD}} $, we adopt Eq. (3) to obtain its concurrence result, i.e., 
\begin{align}
C\left( {{\rho _{{\rm{BPD}}}}} \right) &= \left| {\sqrt {{\lambda _1}\left( {{\rho _{{\rm{BPD}}}}} \right)}  - \sqrt {{\lambda _2}\left( {{\rho _{{\rm{BPD}}}}} \right)} } \right|  
\nonumber \\&= \sqrt {1{\rm{ - }}\eta } C\left( {\left| {{\varphi _{\rm{B}}}} \right\rangle } \right).
\end{align}
And the purity of state ${\rho_{\rm{BPD}}}$ can be given by
\begin{align}
{\rm{Tr}}\left( {\rho _{{\rm{BPD}}}^2} \right) &= 1 - \frac{\eta }{2}{C^2}\left( {\left| {{\varphi _{\rm{B}}}} \right\rangle } \right) \nonumber \\&= 1 - \frac{{{C^2}\left( {\left| {{\varphi _{\rm{B}}}} \right\rangle } \right) - {C^2}\left( {{\rho _{{\rm{BPD}}}}} \right)}}{2}.
\end{align}
Besides, the correlation matrix ${T(\rho_{\rm{BPD}})}$ can be written as
\begin{align}
{T(\rho_{\rm{BPD}}) } &= \left( {\begin{array}{*{20}{c}}
	\sqrt {1{\rm{ - }}\eta } C\left( {\left| {{\varphi _{\rm{B}}}} \right\rangle } \right)&0&0\\
	0&-\sqrt {1{\rm{ - }}\eta } C\left( {\left| {{\varphi _{\rm{B}}}} \right\rangle } \right)&0\\
	0&0&1
	\end{array}} \right)
\nonumber \\ &= \left( {\begin{array}{*{20}{c}}
	C\left(\rho_{\rm{BPD}} \right)&0&0\\
	0&-C\left(\rho_{\rm{BPD}} \right)&0\\
	0&0&1
	\end{array}} \right).
\end{align}
One obtain the quantity $F\left(\rho_{\rm{BPD}} \right)$ of state $\rho_{\rm{BPD}}$ can be expressed as
\begin{align}
F\left( \rho_{\rm{BPD}}  \right) = \sqrt {1+2{C^2}\left( {{\rho _{{\rm{BPD}}}}} \right)}.
\end{align}
Therefore, the relation between steerability $S\left( \rho_{\rm{BPD}}  \right)$ and concurrence $C\left( \rho_{\rm{BPD}}  \right)$ has the following form
\begin{align}
S\left( \rho_{\rm{BPD}}  \right)=C\left(\rho_{\rm{BPD}} \right).
\end{align}
It is apparent that steerability is equivalent to concurrence for the states which is formed by particle (A or B) of Bell-like state going through the PD channel. According to Eq. (23), steerability of the state $\rho_{\rm{BPD}}$ can also be expressed in terms of its purity, i.e., $S\left( {{\rho _{{\rm{BPD}}}}} \right) = \sqrt {{C^2}\left( {\left| {{\varphi _{\rm{B}}}} \right\rangle } \right) + 2{\rm{Tr}}\left( {\rho _{{\rm{BPD}}}^2} \right) - 2}$.

\subsection{Randomly generated two-qubit states}
In Sec. III A and B, we discuss two types of states $\rho_{\rm{BAD}} $ and $\rho_{\rm{BPD}} $, respectively. And the states  $\rho_{\rm{BAD}} $ and $\rho_{\rm{BPD}} $ show the special relation between concurrence and steerability  in Eqs. (18) and (26), respectively. For any two-qubit state, what relation could we obtain about concurrence and steerability?

\textit{Theorem 1.} For a general two-qubit state $\rho $, as long as the sum of concurrence square and purity is greater than 1, then the quantum state $\rho $ exists steerability. And it can be described by the following inequality
\begin{align}
S\left( \rho  \right) \ge \sqrt {\max \left\{ {0,{C^2}\left( \rho  \right) + {\rm{Tr}}\left( {{\rho ^2}} \right) - 1} \right\}}.
\end{align}

\textit{Proof of theorem 1.} The Ref. \cite{w27} gives a complementarity equation of first-order coherence and correlation for a general state $\rho $, 
\begin{align}
\frac{{1 + {D^2}\left( {{\rho _{\rm{A}}}} \right) + {D^2}\left( {{\rho _{\rm{B}}}} \right) + {F^2}\left( \rho  \right)}}{4} = {\rm{Tr}}\left( {{\rho ^2}} \right).
\end{align}
And the Ref. \cite{A} gives a complementarity relation of first-order coherence and concurrence for a general state $\rho $, i.e., 
\begin{align}
\frac{{{D^2}\left( {{\rho _{\rm{A}}}} \right) + {D^2}\left( {{\rho _{\rm{B}}}} \right)}}{2}{{ + }}{{{C}}^2}\left( \rho  \right) \le {\rm{Tr}}\left( {{\rho ^2}} \right).
\end{align}
Combining Eqs. (28) and (29), we can require 
\begin{align}
{F^2}\left( \rho  \right) \ge 2\left[ {{\rm Tr}\left( {{\rho ^2}} \right) + {C^2}\left( \rho  \right)} \right] - 1.
\end{align}
Therefore, based on Eqs. (7) and (30), one derive Eq. (27). By using lots of randomly generated two-qubit states, we visually display the lower boundary of steerability about certain concurrence and purity in Fig.1.

\textit{Theorem 2.} For a general two-qubit state $\rho $, the steerability is bounded  by a quantity related to concurrence and purity. And \textit{theorem 2} can be described by the following inequality
\begin{align}
S\left( \rho  \right) \le \min \left\{ {C\left( \rho  \right),\sqrt {\max \left\{ {0,2{\rm{Tr}}\left( {{\rho ^{\rm{2}}}} \right) - 1} \right\}} } \right\}.
\end{align}
\textit{Proof of theorem 2.} According to Eq. (28), we obtain the relation between the quantity ${F}\left( \rho  \right)$ and purity $\rm{Tr}\left( {{\rho ^2}} \right)$, i.e.,
\begin{align}
{F^2}\left( \rho  \right) \le 4{\rm{Tr}}\left( {{\rho ^{\rm{2}}}} \right) - 1.
\end{align}
If and only if the state $\rho$ belongs to T states \cite{B4,B1,w27} which do not have first-order coherence, Eq. (32) takes the equal sign. Therefore, combining Eqs. (7) and (32), we find that  steerability must be less than the amount related to purity, i.e., $S\left( \rho  \right) \le \sqrt {\max \left\{ {0,2{\rm{Tr}}\left( {{\rho ^{\rm{2}}}} \right) - 1} \right\}}$. It is well known that steering is a sufficient form of quantum inseparabilities. And for pure states, steerability is equal to concurrence as shown in Eq. (11). Therefore, steerability is less than or equal to concurrence, viz., $S\left( \rho  \right) \le C\left( \rho  \right)$. By using lots of randomly generated two-qubit states, we visually display the upper boundary of steerability about certain concurrence in Fig.2.

\begin{figure}
	\centering
	\includegraphics[width=8.2cm]{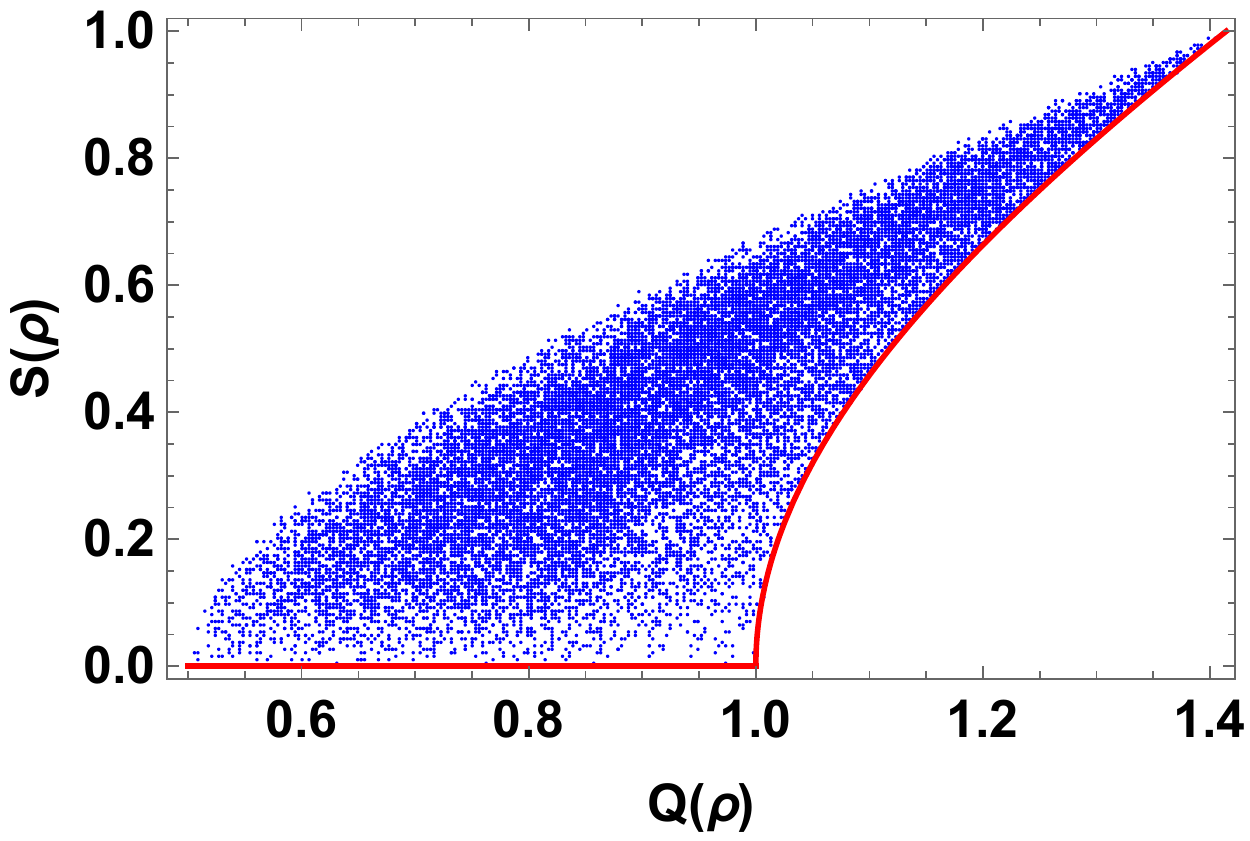}
	\caption{(Color online) Steerability  $S\left( \rho  \right)$ versus the quantity  $Q\left( \rho  \right)$ for two-qubit states $\rho $, where $Q\left( \rho  \right) = \sqrt {{C^2}\left( \rho  \right) + {\rm{Tr}}\left( {{\rho ^2}} \right)} $. The lower bound (red solid line) denotes  $S\left( \rho \right) = \sqrt {\max \left\{ {0,{Q^2}\left( {{\rho }} \right) - 1} \right\}}$, which corresponds to the state $\rho_{\rm{BAD}} $. The figure plots  steerability  $S\left( \rho  \right)$, along the Y axis, and the quantity $Q\left( \rho  \right)$, along the X axis, for $10^5$ randomly generated two-qubit states, by using a specific Mathematica package.}
	\label{Fig1}
\end{figure}

\begin{figure}
	\centering
	\includegraphics[width=8.2cm]{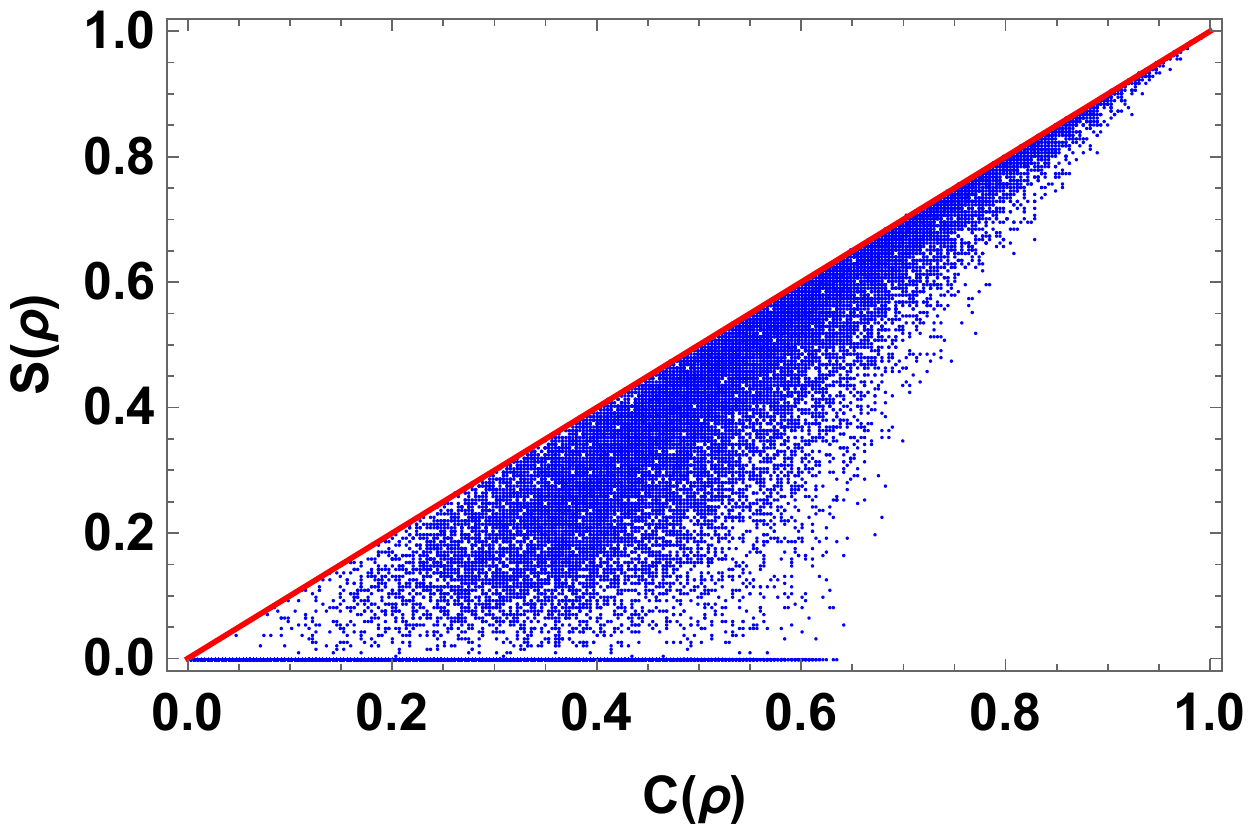}
	\caption{(Color online) Steerability  $S\left( \rho  \right)$ versus concurrence $C\left( \rho  \right)$ for two-qubit  states $\rho $. The upper bound (red solid line) denotes $S\left( \rho  \right) = C\left( \rho  \right)$, which corresponds to the mixed state $\rho_{\rm{BPD}} $ and pure states ${\left| \varphi  \right\rangle }$. The figure plots steerability $S\left( \rho  \right)$, along the Y axis, and concurrence $C\left( \rho  \right)$, along the X axis, for $10^5$ randomly generated two-qubit states, by using a specific Mathematica package.}
	\label{Fig1}
\end{figure}

\section{Steerability and concurrence for a kind of mixed state}
At the front, we have proposed the inequality relation between concurrence and steerability for a general two-qubit state. And we have also investigated the upper and lower upper boundary states. Next, we will study concurrence and steerability for the mixed state $\rho_{\rm WU}$ which is  transformed by performing an arbitrary unitary operation $U$ on Werner-like state ${\rho _{\rm{W}}} = p\left| {{\varphi _B}} \right\rangle \left\langle {{\varphi _B}} \right| +(1 - p)\frac{\mathds{1} \otimes \mathds{1}}{4}$, where $\left| {{\varphi _{\rm B}}} \right\rangle$ is a Bell-like state. And the purity of states $\rho_{\rm WU}$ can be given by ${\rm{Tr}}\left( {\rho _{{\rm{WU}}}^2} \right) ={\rm{Tr}}\left( {\rho _{{\rm{W}}}^2} \right)= \frac{{1 + 3{p^2}}}{4}$.  And the mixed state $\rho_{\rm WU}$ has the following form
\begin{align}
\rho_{\rm WU} = U{\rho _{\rm W}}{U^\dag } = p\left| \varphi  \right\rangle \left\langle \varphi  \right| + (1 - p)\frac{\mathds{1} \otimes \mathds{1}}{4},
\end{align}
where  $\left| \varphi  \right\rangle =U\left| {{\varphi _{\rm B}}} \right\rangle $ is a pure state.
For the state $\rho_{\rm WU}$, we obtain two properties about concurrence $C\left( {{\rho _{\rm WU}}} \right)$ and steerability $S\left( {{\rho _{\rm WU}}} \right)$.

\textit{Property 1.} Steerability $S\left( {{\rho _{\rm WU}}} \right)$ of the state $\rho_{\rm WU}$ is related to steerability ${S}\left( {\left| \varphi  \right\rangle } \right)$ of the pure state $\left| \varphi  \right\rangle $. And the correlation can be expressed as
\begin{align}
S\left( {{\rho _{{\rm{WU}}}}} \right) = \sqrt {\frac{1}{2}\max \{ 0,{p^2}\left[ {1 + 2{S^2}\left( {\left| \varphi  \right\rangle } \right)} \right] - 1\} }.
\end{align}

\textit{Proof of property 1.} The correlation function ${T_{ij}}\left( \rho_{\rm WU}  \right)$ corresponding to the state $\rho_{\rm WU}$ can be reduced as
\begin{align}
{T_{ij}}\left( \rho_{\rm WU}  \right) &= {\rm Tr}\left( {\rho_{\rm WU} {\sigma _i} \otimes {\sigma _j}} \right) \nonumber \\&= p {\rm Tr}\left( {\left| \varphi  \right\rangle \left\langle \varphi  \right|{\sigma _i} \otimes {\sigma _j}} \right) + \frac{{1 - p}}{4} {\rm Tr}\left( {{\sigma _i} \otimes {\sigma _j}} \right) \nonumber \\&= p{T_{ij}}\left( {\left| \varphi  \right\rangle } \right) + \frac{{1 - p}}{4} {\rm Tr}\left( {{\sigma _i}} \right) {\rm Tr}\left( {{\sigma _j}} \right) \nonumber \\&= p{T_{ij}}\left( {\left| \varphi  \right\rangle } \right).
\end{align}
Thus, the value ${F}\left( \rho_{\rm WU}  \right)$ of  state $\rho_{\rm WU} $ is closely related to the value ${F}\left( {\left| \varphi  \right\rangle } \right)$ of the pure state $\left| \varphi  \right\rangle $
\begin{align}
{F}\left( \rho_{\rm WU} \right) = p{F}\left( {\left| \varphi  \right\rangle } \right).
\end{align}
Combining  Eqs. (7), (10) and (11), we can obtain Eq. (34).

\begin{figure}
	\centering
	\includegraphics[width=8.2cm]{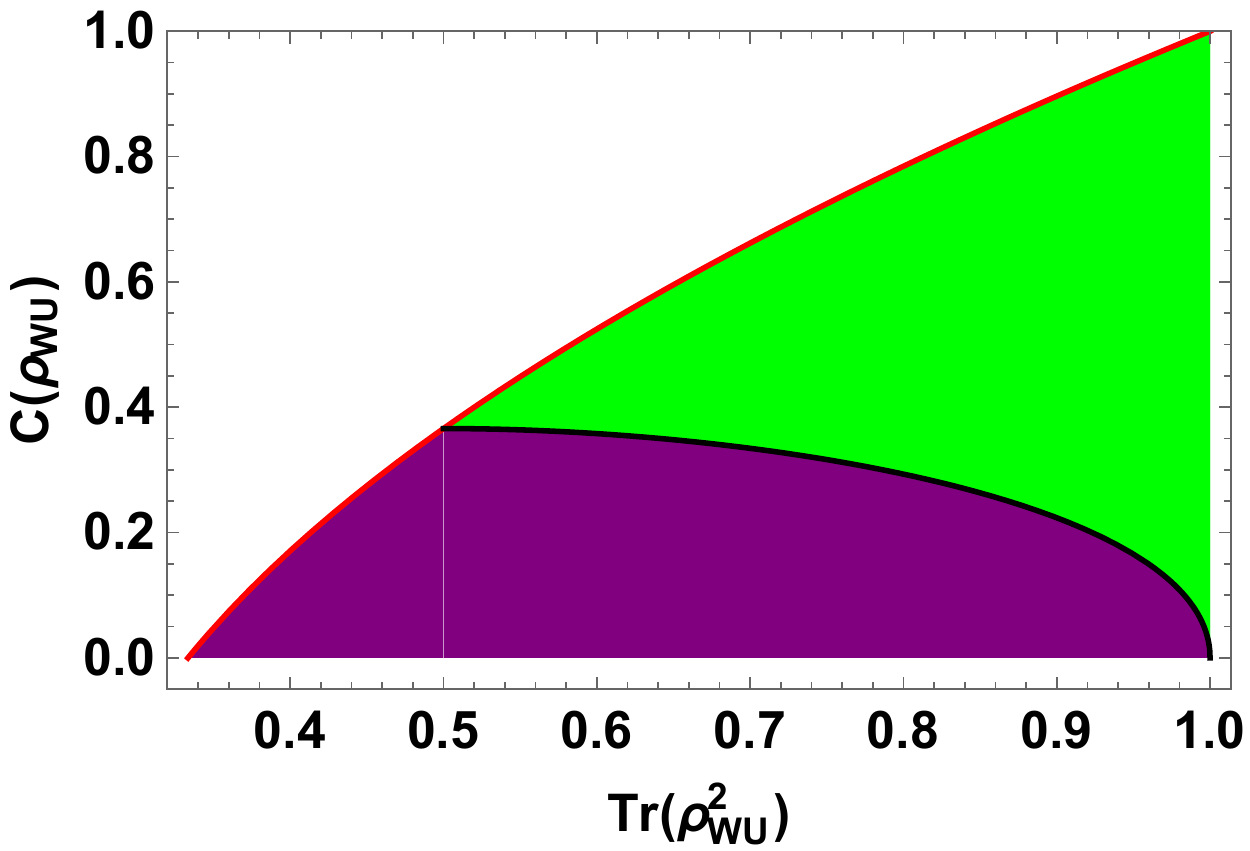}
	\caption{(Color online) Concurrence  $C\left( {{\rho _{\rm WU}}} \right)$ versus purity ${\rm{Tr}}\left( {\rho _{{\rm{WU}}}^2} \right)$ for the state $\rho_{\rm WU}$. The upper bound state (red solid line) corresponds to Werner state. And the solid black curve stands for the sufficient criterion in Eq. (43). The state $\rho_{\rm WU}$ is steerable in the green region. What happens in the purple region is an interesting open question.}
	\label{Fig1}
\end{figure}

\textit{Property 2.} Concurrence $C\left( {{\rho _{\rm WU}}} \right)$ of the state $\rho_{\rm WU}$ is related to concurrence ${C}\left( {\left| \varphi  \right\rangle } \right)$ of the pure state $\left| \varphi  \right\rangle $. And this relation can be expressed as
\begin{align}
C\left( \rho_{\rm WU}  \right) = \max \{ 0,{\rm{ }}pC\left( {\left| \varphi  \right\rangle } \right) - \frac{{1 - p}}{2}\}.
\end{align}

\textit{Proof of property 2.} The non-Hermitian matrix ${\rho _{\rm WU}}{\tilde \rho _{\rm WU}}$ can be given by
\begin{align}
{\rho _{\rm WU}}{\tilde \rho _{\rm WU}} = H + \frac{{{{\left( {1 - p} \right)}^2}}}{{16}} \mathds{1} \otimes \mathds{1},
\end{align}
where $H = {p^2}\left\langle {\varphi } \mathrel{\left | {\vphantom {\varphi  {\tilde \varphi }}}
	\right. \kern-\nulldelimiterspace} {{\tilde \varphi }} \right\rangle \left| \varphi  \right\rangle \left\langle {\tilde \varphi } \right| + \frac{{p\left( {1 - p} \right)}}{4}\left( {\left| \varphi  \right\rangle \left\langle \varphi  \right| + \left| {\tilde \varphi } \right\rangle \left\langle {\tilde \varphi } \right|} \right)$.
Here, since its eigenvalue-equation are too complicated, we do not directly calculate the eigenvalues of non-Hermitian matrix ${\rho _{\rm WU}}{\tilde \rho _{\rm WU}}$. According to some properties of matrix-rank, the rank $R\left( H \right)$ of the matrix $H$ is related to the ranks $R\left( {\left| \varphi  \right\rangle } \right)$ and $R\left( {\left| {\tilde \varphi } \right\rangle } \right)$. And the relation can be given by 
\begin{align}
R\left( H \right) \le& R\left( {\frac{p}{2}\left\langle {\varphi }
	\mathrel{\left | {\vphantom {\varphi  {\tilde \varphi }}}
		\right. \kern-\nulldelimiterspace}
	{{\tilde \varphi }} \right\rangle \left| \varphi  \right\rangle \left\langle {\tilde \varphi } \right| + \frac{{1 - p}}{4}\left| \varphi  \right\rangle \left\langle \varphi  \right|} \right) \nonumber \\ & + R\left( {\frac{p}{2}\left\langle {\varphi }
	\mathrel{\left | {\vphantom {\varphi  {\tilde \varphi }}}
		\right. \kern-\nulldelimiterspace}
	{{\tilde \varphi }} \right\rangle \left| \varphi  \right\rangle \left\langle {\tilde \varphi } \right| + \frac{{1 - p}}{4}\left| {\tilde \varphi } \right\rangle \left\langle {\tilde \varphi } \right|} \right) \nonumber \\ \le & R\left( {\left| \varphi  \right\rangle } \right) + R\left( {\left| {\tilde \varphi } \right\rangle } \right)=2.
\end{align}
It shows that at least two of the eigenvalues of the matrix $H$ are 0. Therefore, two eigenvalues ($\lambda _3$ and $\lambda _4$) of the non-Hermitian matrix ${\rho _{\rm WU}}{\tilde \rho _{\rm WU}}$ can be given by
\begin{align}
{\lambda _3} = {\lambda _4} = \frac{{{{\left( {1 - p} \right)}^2}}}{{16}}.
\end{align}
According to the equations $\sum\nolimits_{n = 1}^4 {{\lambda _n}}  = \rm Tr\left( {\rho _{\rm WU} \tilde \rho _{\rm WU} } \right)$ and $\prod _{n = 1}^4{\lambda _n} =\rm Det\left( {\rho _{\rm WU} \tilde \rho _{\rm WU} } \right)$, we obtain that the other two eigenvalues ($\lambda _1$ and $\lambda _2$) satisfy the following equations
\begin{align}
{\lambda _1} + {\lambda _2} &= {p^2}{C^2}\left( {\left| \varphi  \right\rangle } \right) + \frac{{\left( {1 + 3p} \right)\left( {1 - p} \right)}}{8}, \nonumber \\
{\lambda _1}{\lambda _2} &= {\left[ {\frac{{\left( {1 + 3p} \right)\left( {1 - p} \right)}}{{16}}} \right]^2}.
\end{align}
Combining  Eqs. (3), (40) and (41),  concurrence $C\left( {{\rho _{\rm WU}}} \right)$ of the states $\rho _{\rm WU}$ can be expressed as
\begin{align}
C\left( {{\rho _{\rm WU}}} \right) &= \max \left\{ {0,\sqrt {{\lambda _1}}  - \sqrt {{\lambda _2}}  - \sqrt {{\lambda _3}}  - \sqrt {{\lambda _4}} } \right\} \nonumber \\ &= \max \{ 0,\sqrt {{\lambda _1} + {\lambda _2} - 2\sqrt {{\lambda _1}{\lambda _2}} }  - \frac{{1 - p}}{2}\}  \nonumber \\ &= \max \left\{ {0,pC\left( {\left| \varphi  \right\rangle } \right) - \frac{{1 - p}}{2}} \right\}.
\end{align}
Obviously, when the pure state ${\left| \varphi  \right\rangle }$ is a Bell state, both concurrence and steerability of the states  $\rho _{\rm WU}$ reach maximum. Eq. (34) shows that CJWR inequality can be violated at the case of $p > \frac{1}{{\sqrt {1 + 2{C^2}\left( {\left| \varphi  \right\rangle } \right)} }}$ for the state $\rho _{\rm WU}$. Conbining Eqs. (11), (34) and (37),  steerability can be obtained by concurrence and purity, i.e., 
\begin{align}
S\left( {{\rho _{{\rm{WU}}}}} \right) = \sqrt {\max \left\{ {0,x\left( {{\rho _{{\rm{WU}}}}} \right) + Q^2\left( {{\rho _{{\rm{WU}}}}} \right)}-1 \right\}},
\end{align}
where $x\left( {{\rho _{{\rm{WU}}}}} \right) = \frac{{1 + 2C\left( {{\rho _{{\rm{WU}}}}} \right)}}{2}\left( {1 - \sqrt {\frac{{4{\rm{Tr}}\left( {\rho _{{\rm{WU}}}^2} \right) - 1}}{3}} } \right)$ and $Q\left( {{\rho _{{\rm{WU}}}}} \right) = \sqrt {{C^2}\left( {{\rho _{{\rm{WU}}}}} \right) + {\rm{Tr}}\left( {\rho _{{\rm{WU}}}^2} \right)}$.
Conbining  \textit{Theorem 2}, we can obtain a sufficient criterion for detection steering by using concurrence and purity for the state $\rho _{\rm WU}$ (as shown in Fig. 3).

\section{Conclusion} \label{sec6}
In this paper, we have investigated the constraint relation between steerability and concurrence for two kinds of evolutionary states and lots of randomly generated two-qubit states. The result shows that the upper and lower boundaries of steerability for any two-qubit state can be exactly expressed based on certain concurrence and purity. Specifically, the lower boundary can be used as a sufficient criterion for steering detection. In other words,  a general two-qubit state must be steerable if the sum of purity and concurrence's square is greater than one.  And the upper boundary reveals that steerable states form a strict subset of entangled states. Futhermore, we consider a special kind of mixed state transformed by performing an arbitrary unitary operation on Werner-like state. It can be obtained  that the special mixed state’s concurrence and steerability are related to ones of a pure state transformed by the unitary operation performed on Bell-like state, respectively. And we  present and demonstrate a sufficient criterion, which provides an effective theoretic basis to seek steerable states from entangled states.

\section*{Acknowledgements} 
This work was supported by the National Science Foundation of China under Grant Nos. 11575001 and 61601002, Anhui Provincial Natural Science Foundation (Grant No. 1508085QF139) and Natural Science Foundation of Education Department of Anhui Province (Grant No. KJ2016SD49).

\bibliographystyle{plain}

\end{document}